\documentclass[aps,prl,twocolumn,floats,showpacs,superscriptaddress]{revtex4}
\usepackage{graphicx,epsfig}
\usepackage{times}
\usepackage{graphics,dcolumn,bm,float}
\usepackage{amssymb,amsmath,rotate,color}
\usepackage{psfrag}

\begin{document}
\unitlength 1 cm
\newcommand{\be}{\begin{equation}}
\newcommand{\ee}{\end{equation}}
\newcommand{\bearr}{\begin{eqnarray}}
\newcommand{\eearr}{\end{eqnarray}}
\newcommand{\nn}{\nonumber}
\newcommand{\vk}{\vec k}
\newcommand{\vp}{\vec p}
\newcommand{\vq}{\vec q}
\newcommand{\vkp}{\vec {k'}}
\newcommand{\vpp}{\vec {p'}}
\newcommand{\vqp}{\vec {q'}}
\newcommand{\bk}{{\mathbf k}}
\newcommand{\bp}{{\mathbf p}}
\newcommand{\bq}{{\mathbf q}}
\newcommand{\br}{{\mathbf r}}
\newcommand{\bR}{{\mathbf R}}
\newcommand{\bd}{{\mathbf d}}
\newcommand{\up}{\uparrow}
\newcommand{\down}{\downarrow}
\newcommand{\fns}{\footnotesize}
\newcommand{\ns}{\normalsize}
\newcommand{\cdag}{c^{\dagger}}

\title{Dynamical Mean Field Study of The Dirac Liquid}

\author{S. A. Jafari{\footnote {Electronic address:
sa.jafari@cc.iut.ac.ir}}}

\affiliation{Department of Physics, Isfahan University of
Technology, Isfahan 84154-83111, Iran}
\affiliation{The Abdus Salam ICTP, 34100 Trieste, Italy}

\begin{abstract}
Renormalization is one of the basic notions of condensed matter physics.
Based on the concept of renormalization, the Landau's {\em Fermi liquid}
theory has been able to explain, why despite the presence of Coulomb interactions,
the free electron theory works so  well for simple metals with extended Fermi surface (FS).
The recent synthesis of graphene has provided the condensed matter physicists
with a low energy laboratory of Dirac fermions where instead of a FS, one has two Fermi points. 
Many exciting phenomena in graphene
can be successfully interpreted in terms of free Dirac electrons. 
In this paper, employing dynamical mean field theory (DMFT), 
we show that an interacting Dirac sea is essentially an 
effective free Dirac theory. This observation suggests the notion of {\em Dirac liquid} 
as a  fixed point of interacting 2+1 dimensional Dirac fermions.
We find one more fixed point at strong interactions describing a Mott insulating state,
and address the nature of semi-metal to insulator (SMIT) transition in this system.
\end{abstract}

\maketitle

\section{introduction}

  Dirac theory of electrons was formulated in 1928 to describe the relativistic 
motion of electron waves~\cite{WeinbergBook}. 
This theory not only is consistent with the spin $\hbar/2$ of  electron, but also obeys
the correct relativistic covariance. Now days, accelerators and neutron stars are not 
the only places to search for Dirac fermions. Advances in science and technology has enabled
physicists to realize Dirac fermions in energy scales as low as $1eV$, or even lower 
in solid state physics. The nodal qasiparticles of d-wave cuprate superconductors 
being one example in superconducting state of matter~\cite{Tsuei}.

Recently, graphene, a single atomic layer of graphite, 
was fabricated~\cite{Novoselov}, which has initiated a rapidly growing 
research activity in condensed matter physics~\cite{NetoRMP}.
The low-energy electronic structure of this system
is approximately described by a 2+1 dimensional Dirac theory~\cite{Semenoff},
which enjoys a chiral symmetry~\cite{Semenoff,KatsnelsonChiral}. 
Therefore graphene, the unrolled carbon nano-tube; provides
the condensed matter with a laboratory for the relativistic fermions 
on the table top~\cite{Vinu}.

  There is still another road to search for Dirac fermions in material physics laboratory.
Recent advances in ultra cooling and atom trap 
methods~\cite{BEC} have elevated this technology to a chip based level~\cite{Aubin}.
This provides a unique opportunity for tuning the interaction parameters
in microscopic models employing the so called Feshbach resonance~\cite{Feshbach}.
These motivate the study of strongly interacting fermions on honeycomb lattice 
in parameters regimes, much beyond what can be currently realized in graphene,
or high $T_c$ cuprates.
\begin{figure}[t]
   \begin{center}
   \vspace{-1.0cm}
   \includegraphics[width=6cm,angle=-00]{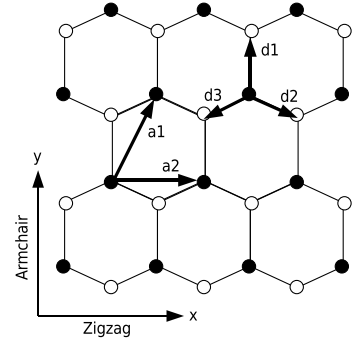}
   \vspace{-1.5cm}
   \caption{Lattice structure of a honeycomb lattice as a superposition of two simple
   monoclinic Bravais lattices. The C-C distance is $a\approx 1.42\AA$, such that the lattice constant
   becomes $\sqrt 3 a$ and the basis vectors
   are given by ${\mathbf a}_1= 3a/2\hat {\mathbf e}_x+\sqrt 3 a/2\hat{\mathbf e}_y$,
   ${\mathbf a}_2=\sqrt 3 a\hat{\mathbf e}_x$.}
   \label{lattice.fig}
   \end{center}
\end{figure}

\begin{figure}[t]
   \begin{center}
   \psfrag{D = 2}{\hspace{-5pt}D=$2$}
   \psfrag{D = infinity}{\hspace{10pt}D=$\infty$}
   \includegraphics[width=5cm,angle=-90]{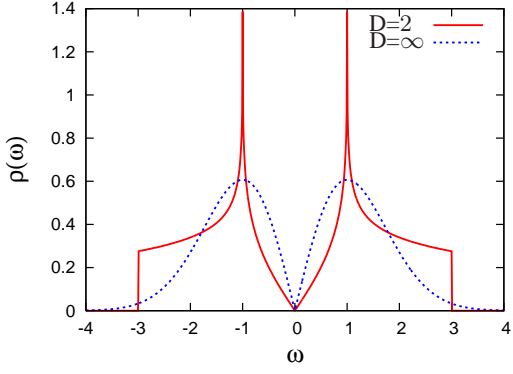}
   \caption{Density of states for tight binding fermions on honeycomb ($D=2$) 
   and hyper honeycomb ($D=\infty$) lattices. In the case of undoped graphene,
   due to spin degeneracy the valence band (mirror image of this figure) is 
   completely filled. For spin-less fermions on optical lattices at quarter filling
   also the chemical potential is at $\omega=0$. Energies are in units of hopping 
   amplitude $t$.}
   \label{DOSs.fig}
   \end{center}
\end{figure}

   The three-fold coordination for hopping of fermions on a honeycomb
lattice is responsible for the relativistic low-energy theory~\cite{Semenoff},
which strictly speaking describes a semi-metallic state of a Dirac sea; rather than 
a metallic state of a Fermi sea. This semi-metal is described by a pseudo-gap
in the density of states (DOS) shown in Fig.~\ref{DOSs.fig}. Such feature in DOS 
is also relevant to the nodal quasi particles of high $T_c$ 
cuprate superconductors~\cite{Tsuei}. 

   While the single particle Dirac theory seems to work well in 
graphene, it is important to search for interesting many-body
effects on honeycomb lattice. For example the possibility of 
$f$-wave or $d+id$-wave superconductivity~\cite{Honerkamp,Doniach,BaskaranMgB2,Shenoy},
as well as CDW instability~\cite{Honerkamp} on honeycomb lattice
has been discussed.
   Perturbative renormalization group studies~\cite{GuineaPRB99,Aleiner} has indicated
that the long range part of the Coulomb interaction is irrelevant, and
gives rises to a non interacting fixed point. In this paper we focus on the
short range part of the interaction, employing a non-perturbative
method of DMFT. This theory has been very successful in addressing
the question of metal to insulator transitions~\cite{RMPDMFT}.
According to the prediction of DMFT, the onset of transition
to Mott insulating phase is accompanied by formation of the so called
Kondo resonance, the spectral weight of which characterizes the
quasiparticle weight $Z$ of the underlying Fermi sea.
In the case of a half-filled Dirac sea, instead of an extended
Fermi surface, one has to deal with two {\em Fermi points} at the so called
$K$ points of the Brillouin zone.
Due to the cone like dispersion of the conduction and valence band near these
points, there will be no quasi particle state at the Fermi level~\cite{NetoRMP}. 
Hence the quasiparticle weight $Z$ of Fermi liquids can not be used
to discuss the transition to Mott insulating regime.

In this work we employ the DMFT approximation
to study the SMIT of {\em Dirac liquids} in two dimensional honeycomb lattice.
The picture which emerges from this study is that; for weak to moderate
interaction strengths, the Dirac sea state 
remains stable against local many body interactions. The sole
role of interactions would be to {\em renormalize} the Dirac quasi
particles. Such robustness of the Dirac fermions against many body
interactions has been observed in various measurements in 
graphene~\cite{Rotenberg}.
For strong enough interactions, a Mott insulating state is stabilized.
We study some simpler model density of states which mimic the true
2D DOS of graphene Eq.~(\ref{dos2.eqn}). We find the value of $U_c \sim 13.3 t$,
which is not sensitive to the details of DOS. 

 \section{Model and method}
We take the tight-binding electrons on honeycomb lattice~\cite{NetoRMP},
which give rise to Dirac spectrum near the $K$ points of the
Brillouin Zone (BZ), and add a short ranged Coulomb 
interaction of Hubbard type to it:
\be
   H=-t\sum_{\langle i,j\rangle\sigma}
   c^\dagger_{i\sigma}c_{j\sigma}+\mbox{h.c.}
   +U\sum_j \left(n_{j\up}-\frac{1}{2}\right)
   \left(n_{j\down}-\frac{1}{2}\right)
\ee
where the hoping amplitude for $p_z$ electrons in graphene is $t\sim 2.7 eV$,
while for the atoms trapped in optical lattices it is typically on the scale
of $t\sim \mu K\sim 10^{-10} eV$.
Through the paper when the units are not specified, the units in which 
$t=\hbar=1$ is implied. Also we take the atomic separation to be $1$.
This model is manifestly particle-hole symmetric and the Hartree term at half 
filling ($\langle n_{j\sigma}\rangle=1/2$) vanishes. 

In the absence of interaction, the Hamiltonian becomes
\be
   H_0=\left(
   \begin{array}{cc}
      0 & -f(\bk)\\
      -f^*(\bk) & 0
   \end{array}\right)
   \label{H0.eqn}
\ee
where $\bk=(k_x,k_y)$, and $f(\bk)=e^{i\bk.\bd_1}+e^{i\bk.\bd_2}+e^{i\bk.\bd_3}$ 
(Fig.~\ref{lattice.fig}). The eigenvalues of this matrix are $\varepsilon_\bk=|f(\bk)|$.
When the above complex function
$f(\bk)$ linearized around the Fermi points $K,K'$ of the BZ,
gives rise to the Hamiltonian $H=\pm v_F{\mathbf \sigma}.\bk$.
The DOS for this non interacting system is given by~\cite{Hobson},
\be
   \rho(\varepsilon)=\frac{|\varepsilon|}{\pi^2} \frac{1}{\sqrt{Z_0}}
   F\left(\frac{\pi}{2},\sqrt{\frac{Z_1}{Z_0}} \right),
   \label{dos2.eqn}
\ee
where
\be
   Z_0=\left\{
   \begin{array}{lr}
      \left(1+|\varepsilon|\right)^2-\left(\varepsilon^2-1\right)^2/4; & |\varepsilon|<1\\
      4|\varepsilon|;					& 1\le |\varepsilon|\le 3
   \end{array}
   \right.,
\ee
and 
\be
   Z_1=\left\{
   \begin{array}{lr}
      4|\varepsilon|;	& |\varepsilon|<1\\
      \left(1+|\varepsilon|\right)^2-\left(\varepsilon^2-1\right)^2/4; & 1\le |\varepsilon|\le 3\\
      \end{array}
   \right..
\ee
Here $F(\pi/2,x)$ is the complete elliptic integral of first 
kind~\cite{Abramowitz}. 
At low energies where the dispersion becomes $\varepsilon_\bk=\pm v_F k$, 
we have linear energy dependence in the pseudo-gap shaped DOS (Fig.~\ref{DOSs.fig}),
\be
   \rho(\varepsilon)=2\pi v_F^{-2}\left|\varepsilon \right|
   \label{lineardos.eqn}
\ee
where $\hbar v_F= 3ta/2$ is the bare Fermi velocity of quasiparticles 
at the Fermi points.
In Fig.~\ref{DOSs.fig} we have plotted the DOS for two and
infinite dimensional (hyper)honeycomb lattice~\cite{Santoro}.

  In DMFT, one starts with a local free propagator
$g_{0\alpha}(\omega)$, where the Greek indices $\alpha,\beta$, etc.
correspond to sub-lattices A,B. Next and most important part of
the DMFT consists in employing an impurity solver to obtain the
local self-energy $\Sigma_{\alpha}$ as a functional of $g_{0\alpha}$. 
Note that, within the single-site DMFT approximation the
self-energy is purely local~\cite{RMPDMFT} which means the off-diagonal
self-energies vanish and it can be described
by the diagonal matrix elements $\Sigma_\alpha(\omega)$.
We use the iterated perturbation theory to solve the impurity problem: 
\be
   \Sigma_\alpha(t)=U^2g_{0\alpha}^2(t) g_{0\alpha}(-t).
   \label{sopt.eqn}
\ee
Now using this self-energy we construct the Green's function 
of the honeycomb lattice which can be written as 
\be
   G_{\alpha\beta}(\bk,\omega)=\left(
   \begin{array}{cc}
      \zeta_A(\bk,\omega) & -f(\bk)\\
      -f^*(\bk) & \zeta_B(\bk,\omega)
   \end{array}\right)^{-1},
   \label{gkwhoney.eqn}
\ee
where $\bk$ belong to the first  BZ of of sub-lattice, and  
$\zeta_\alpha(\bk,\omega)=\omega-\varepsilon_\bk+\mu-\Sigma_\alpha(\omega)$.
The half filled band which gives rise to a Dirac sea, due to the 
bipartite nature of the honeycomb lattice corresponds to $\mu=0$.
Such half-filling condition corresponds to undoped graphene, whereas in
optical lattices with one specie atom, it corresponds to half atom
per site. Note that the particle-hole symmetry can also be used
which implies $\Sigma_A(\omega)=-\Sigma_B(-\omega)$.
To connect the lattice Green's function to local one, we need to 
project~(\ref{gkwhoney.eqn}) onto sub-lattice $A$($B$), which reads
\bearr
   g_\alpha(\omega)=\xi_{\bar\alpha}\sum_\bk \frac{1}{\xi_A\xi_B-\varepsilon_\bk^2}
   = \xi_{\bar\alpha}\int d\varepsilon 
      \frac{\rho(\varepsilon)}{\xi_A\xi_B-\varepsilon^2}
   \label{projectedG.eqn}
\eearr
where $\rho(\varepsilon)$ is the DOS of massless Dirac fermions, Eq.~(\ref{dos2.eqn}).
To close the set of equations, we only need to append the Dyson equation
\be
   g_\alpha^{-1}(\omega)=g_{0\alpha}^{-1}(\omega)-\Sigma_\alpha(\omega)
   \label{dyson.eqn}
\ee
We solve the set of equations (\ref{sopt.eqn}), (\ref{gkwhoney.eqn}),
(\ref{projectedG.eqn}) and (\ref{dyson.eqn}) self-consistently.

\section{Results}
\begin{figure}[t]
  \begin{center}
    \includegraphics[width=6cm,angle=-90]{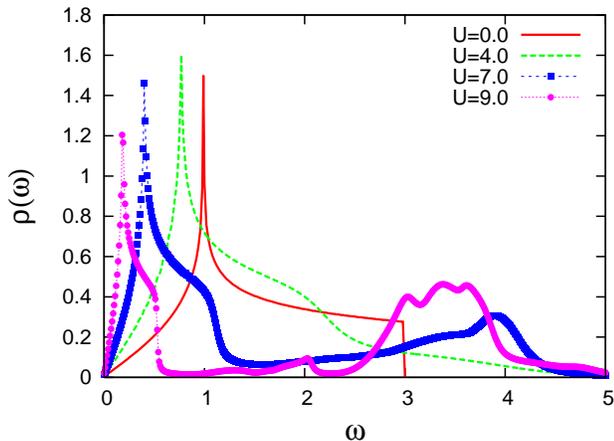}
    \caption{(Color online) Changes in the band structure of Dirac
    Fermions on honeycomb lattice as a function of Hubbard $U$. 
    For values of $U$ larger than the bandwidth $W=6t$, the upper
    and lower (not shown in this figure) Hubbard bands set in.} 
    \label{graphene-dos.fig}
  \end{center}
\end{figure}
The Hilbert transformation in equation~(\ref{projectedG.eqn}) is the only 
place where the band structure (DOS) of system enters the DMFT machinery.
Using the DOS of realistic graphene given in Eq.~(\ref{dos2.eqn}) we obtain
the interacting DOS for various values of $U<U_c\approx 13.3t$ shown in 
Fig.~\ref{graphene-dos.fig}. As can be seen, upon increasing $U$, the
spectral weight is transferred to higher energies, accompanied by a 
increase in the slope of the DOS. Interpreting the results in terms of 
Eq.~(\ref{lineardos.eqn}) one sees that (i) the Dirac nature of the spectrum 
is preserved, manifested in the linear DOS and the pseudo-gap structure
around the charge neutrality point is preserved for all $U<U_c$,
but with a reduced Fermi velocity $\tilde v_F$ replacing the non-interacting
Fermi velocity $v_F=3/2$. (ii) As there are no quasiparticle states at the 
Fermi level for the Dirac fermions, the SMIT is not accompanied by a Kondo resonance
at the Fermi surface. (iii) The (logarithmic) van Hove singularity of the band 
structure corresponding to saddle point at the M point of the Brillouin zone
persists.

\begin{figure}[t]
  \begin{center}
    \includegraphics[width=6cm,angle=-90]{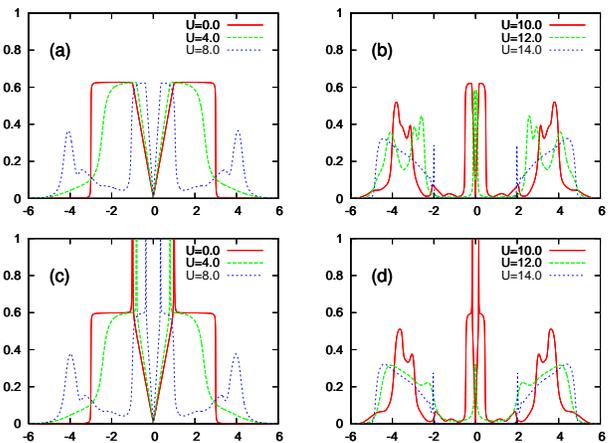}
    \caption{(Color online) DOS versus energy. Energy is in 
    units of hopping parameter $t$.
    (a), (c): Renormalization of Dirac sea by Hubbard
    interaction $U$. (b), (d): SMIT for Dirac fermions. 
    Upper two panels (a,b) correspond to $h=0$, while the
    lower two panels (c,d) correspond to $h=1$ in the toy DOS
    introduced in the text. The values of $U$ are in units of $t$.} 
    \label{dosmodel.fig}
  \end{center}
\end{figure}

  For the DOS~(\ref{dos2.eqn}) the Hilbert transformation must
be done with a numerical quadrature with typically $N_a=2000$ abscissas
for a Simpson quadrature to achieve the double precision accuracy needed
in typical self-consistency problems. We simplify the DOS of 
graphene with a parameter dependent DOS which mimics the essential
features of the graphene band structure, but at the same time
allows for analytic evaluation of the Hilbert transform~(\ref{projectedG.eqn})
which saves a lot of time:
\be
   \rho_h(\varepsilon)=h\delta(|\varepsilon|-1)
   +\left\{ \begin{array}{ll}
   2\pi v_F^{-2}|\varepsilon| & |\varepsilon|<1 \\
   2\pi v_F^{-2}              & 1<|\varepsilon|\le 3\\
   0                          & {\rm otherwise}
   \end{array}\right.
   \label{dosmodel.eqn}
\ee
This model DOS captures the linear dispersion around the charge neutrality
point and a sharp singularity of strength $h$ at $|\varepsilon|=\pm 1$. 
Solving the DMFT equations with DOS~(\ref{dosmodel.eqn}) in the
projection formula~(\ref{projectedG.eqn}) we obtain figure~\ref{dosmodel.fig}.
Panels (a), (b) correspond to $h=0$ and (c), (d) denote $h=1$. 
We see that this model DOS produces the essential features (i)-(iii)
above.
For small to intermediate values of $U$ one can see the renormalization
of the slope in (a), (c). Comparison of (a), (c) shows that this feature
is qualitatively independent of the presence or lack of a singularity.
Singularity places initially at $|\varepsilon|=\pm 1$ moves to lower
energies as one increases $U$. Panels (b), (d) show the SMIT for
$h=0$, $h=1$, respectively. For $h=0$, we obtain $U_c^{\rm h=0}\approx 13.3 t$,
while for $h=1$, the critical value is slightly lower.
Right at the critical point, due to critical slowing down,
it is extremely difficult to obtain convergence. The reason is that
for $U= U_c^-$, the DOS slope diverges (meaning $\tilde v_F\to 0$), and 
the pseudo-gap tends to closes.

\begin{figure}[t]
  \begin{center}
    \includegraphics[width=5.5cm,angle=-90]{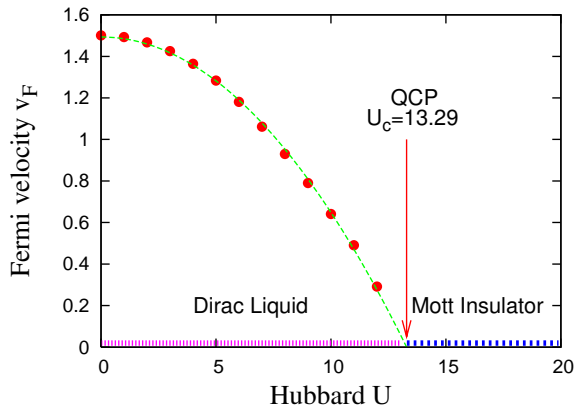}
    \caption{(Color online) Renormalization of the Dirac fermions velocity
    $v_F$ by the Hubbard $U$ term.} 
    \label{vf.fig}
  \end{center}
\end{figure}
In Fig.~\ref{vf.fig} we show the renormalized Fermi velocity $\tilde v_F$ 
for the model DOS corresponding to $h=0$ as a function of $U$.
The Fermi $\tilde v_F$ of Dirac fermions interpolates
to zero at the quantum critical point (QCP) $U_c\approx 13.3t$. Unlike the MIT 
in metallic systems with extended Fermi surface, there is no Kondo
resonance corresponding to quasiparticle at the Fermi level. Therefore
SMIT can not be described in terms of the spectral weight $Z$ of such
resonant state. Instead the $\tilde v_F$ in Fig.~\ref{vf.fig} 
can be identified as an order parameter characterizing a {\em Dirac liquid}
state. Beyond a QCP at $U_c\approx 13.3t$, a Mott insulating state
appears.

The semi-metal to insulator transition in $D=\infty$ honeycomb lattice 
(Dotted line in Fig.~\ref{DOSs.fig}) has been previously studied by Santoro and 
coworkers~\cite{Santoro}, where they focused on the zero and finite temperature
magnetic phase transitions. They find that at zero temperature, at a critical
value of $U_c^\infty\approx 2.3 W$ there is a phase transition from a paramagnetic 
semi-metal to an anti-ferromagnetic Mott insulator, where $W$ is the band width.
If one takes this as a ad hoc critical value for $D=2$ honeycomb lattice with $W=6t$,
one expects a critical value of $U_c^\infty\approx 13.8t$ in 2D honeycomb lattice,
which is very close to the value $13.3t$ we find here.
Sorella and Tosatti approached the same problem by quantum Monte Carlo
methods~\cite{Sorella}. They found a zero temperature SMIT between a
non magnetic semi-metal and an anti-ferromagnetic insulator at a critical
value of $U_c^{\rm QMC}\approx 4.5 t$. The Brinkman-Rice
analysis of the SMIT within the Gutzwiller approximation gives 
a critical value of $U_c^{\rm BR}\approx 12.8t$~\cite{Balatsky},
which is again rather close to the value we obtain.
The corresponding mean field value is $U_c^{\rm MF}\approx 2.23 t$.
Such a significant difference between the mean field and more accurate
methods indicates the importance of quantum fluctuations in SMIT.

\section{Summary and discussion} 
   In this paper we studied the zero temperature phase transition of 
Dirac fermions on a honeycomb lattice. Our results can be applied to
atoms in honeycomb optical lattices, as well as electrons in undoped
graphene. We studied the nature of SMIT in this system. We found no
quasiparticle resonance state. The renormalized Fermi velocity $\tilde v_F$
was suggested as an order parameter for a Dirac liquid separated from
a Mott insulator by a QCP at $U_c\approx 13.3 t$.
The recent renormalization in $v_F$ seen in ARPES measurements~\cite{Rotenberg}
can not be solely described in terms of electron-phonon or electron-plasmon
interactions of a doped graphene. Such renormalization arising from
Hubbard term may also contribute to those seen in ARPES measurements.
 
Others have also reported on a quantum critical point behavior
of Dirac fermions using large $N$ expansion, beyond which an insulating
phase emerges~\cite{SonQCP}.
The renormalization group flow equations for quartic perturbations~\cite{DrutSon}
indicate possible insulating phase at strong couplings.
Foster and Aleiner find logarithmic divergences using large $N$
renormalization group approach~\cite{Aleiner}.
They found that the long range part of the Coulomb interaction enhances
the short range part of the interaction. Our method addresses the role
of short range part in a non-perturbative DMFT sense, which therefore
can be regarded as a complementary analysis to their's.

Gonzalez and coworkers find within renormalization scheme that the Coulomb 
interactions drive the system to a non-interacting fixed point~\cite{GuineaPRB99}.
Their finding agrees with the Dirac Liquid fixed point in our analysis. However
our non-perturbative DMFT analysis indicates another Mott insulating fixed 
point in addition to the Dirac liquid fixed point of Ref.~\cite{GuineaPRB99}
in agreement with our exact and non-perturbative analysis.

In graphene samples $U\approx 6$ eV, which is not strongly correlated, 
places them at $U/t\sim 2.2$, far from the Mott insulating phase.
However, one can describe the graphene starting from a Mott insulating
resonance valence bond (RVB)~\cite{BaskaranJafari,PaulingBook} point.
In such a description one needs to allow for charge fluctuations on 
top of an RVB ground  state~\cite{BaskaranMgB2}. This provides
an opportunity for high temperature superconductivity in $\sim 20\%$ doped 
graphene~\cite{Doniach,Shenoy}.

\section{acknowledgements}
This work was supported by ALAVI Group Ltd.
The author thanks F. Shahbazi, K. Esfarjani and Igor Aleiner for useful discussions,
and the Abdus Salam ICTP for hospitality in a summer visit, during which this
research was done.

\end{document}